\documentclass[a4paper,12pt]{article}
\usepackage{amsmath,amssymb}
\newtheorem{proposition}{Proposition}

\begin{document}

\begin{center}
\section*{The anyon model: an example inspired by string theory}

{S.V. Talalov}

\vspace{5 mm}

\small{Department of Theoretical Physics, State University of Tolyatti,\\
 14 Belorusskaya str., Tolyatti, Samara region, 445667 Russia.\\
svtalalov@tltsu.ru}

\end{center}

\begin{abstract}
{~~~ We investigate the enlarged class of  open finite strings in $(2+1)D$ space-time. 
The new  dynamical system  related to this class is constructed  and quantized here.
As the result, the energy spectrum of the model is defined by a simple formula
${\sf S} = \alpha_n{\sf E} + c_n$;  the spin ${\sf S}$ is an arbitrary number here but  the constants $\alpha_n$  and 
$c_n$ are eigenvalues for
certain spectral problems in fermionic Fock space ${\bf H}_\psi$ constructed for the free $2D$ fermionic field.}
\end{abstract}

{\bf keywords:}  anyon models, noncritical strings, boson-fermion correspondense.

 {PACS numbers:  11.25.Pm, 04.60.Kz, 02.40.Hw.}

\section{Introducton}
 \indent
As it seems, ''anyons''\cite{Wil} --
 the particles with arbitrary spin and statistics -- can be realised as the exitations of some infinity-dimensional 
dynamical system on a plane\cite{Lau}.  The finite planar string is the simplest example of this  system.
Note that a non-standard point of view on
elementary particles  was suggested recently\cite{LevWen}. It described them as the defects of the string condensed matter
(''string-net condensation'').
Open string in arbitrary space-time
dimensions is a well-investigated object (see, e.g. \cite{Zwi,GShW}). One of the frequently discussed structures
here is the first form {\bf I} of the world-sheet -- as opposed to
the second form {\bf II}. In this work we investigate  the
 finite string on a spatial plane in terms of
second form {\bf II} and construct the certain dynamical system related to this string.
We  interprete  the excitations of constructed sysytem as  anyon-type quasiparticles.

  Let us begin with the classical theory.
 The suggested scheme  \cite{TMF00} generalizes  the standard
 geometrical approach \cite{BarNes} in  string theory. Here we will  briefly remind the main points of our approach.
 We   start with the
    Nambu-Goto action
  \begin{equation}
 S = - {\gamma} \int\sqrt{-\det(\partial_i {\bf X} \partial_j {\bf X})}  d\xi^0d\xi^1\,,
 \label{action}
 \end{equation}
  where  ${\bf X} = {\bf X}(\xi^0,\xi^1)$  is the vector in Minkowski space-time $E_{1,3}$, the parameters 
   $\xi^0$ and $\xi^1$ are world-sheet parameters and
symbols  $\partial_i$ denote corresponding derivatives. Thus we will consider the minimal surfaces in the space-time $E_{1,3}$.
It is well-known  (see, for example \cite{Zwi,BarNes,And})
that  the special parametrization of the world-sheet can be selected -- so that both   
  equalities
  \begin{equation}
  \label{eq_x}
\partial_{+}\partial_{-}{\bf X} =0\,,
\end{equation}
 the constraints
  \begin{equation}
  \label{con_x}
\qquad  (\partial_{\pm}{\bf X})^2 = 0\,
\end{equation}
and the boundary conditions
\begin{equation}
  \label{bc_x}
  \partial_{1}{\bf X}\Bigg\vert_{\xi^1=0}= \quad   \partial_{1}{\bf X}\Bigg\vert_{\xi^1=\pi}= \quad 0\,
   \end{equation}
   will be fulfilled. 
 We denote  $\partial_\pm = \partial/\partial \xi_\pm $ and $\xi_\pm =\xi^1\pm\xi^0$ here.
   Thus  the initial objects of our investigation are the time-like world-sheets with orthonormal parametrization.
 Our initial steps in  sections 1 and 2 will be:
 \begin{itemize}
 \item We will reduce our theory to $3D$ case;
    \item  We will define new bijective parametrization for the world-sheet variables ${\bf X}(\xi^0,\xi^1)$ constrained by the 
    equalities (\ref{con_x}) so that
\begin{equation}
\label{param1}
        {\bf X}(\xi^0,\xi^1)  = 
        {\bf X}\,\Bigl[\,\varkappa ;\, {\bf Z},\,  {B};\, \varphi(\xi^0,\xi^1),\, \alpha_+(\xi^0,\xi^1),\, \alpha_-(\xi^0,\xi^1)\Bigr]\,,
\end{equation}
   where the new (unconstrained) parameters will be transformed differently for the scale and Poincar\'e transformations of space-time. So,
   the Poincar\'e transformations of the world-sheet will transform the constant vector ${\bf Z}\in E_{1,2}$ and the constant 
   matrix $B\in SL(2,R)$ only; the scale transformations will transform the real constant $\varkappa\in (0,\infty)$ only. The 
   functions  $\varphi$,  $\alpha_\pm$ will be certain relativistic and scale invariant  functions. 
   \item We will factorize the set ${\mathcal X}$ of the considered world-sheets on the orbits for certain gauge group
   ${\sf G}_0$ . 
               \end{itemize}

    In  section 3 we consider  the set   ${\mathcal X}/{\sf G}_0$  as a dynamical system; the $\xi^0$-dynamics is defined here by the
    differential equations (\ref{eq_x}) and the conditions (\ref{con_x}) and (\ref{bc_x}).  The well-known fact that same dynamical system can have the
    different hamiltonian structures (see\cite{MSSV}, for example).  In accordance   with the Dirac ideas\cite{Dirac}  we define the 
    hamiltonian structure     for our dynamical system  as initial conception.  The constructed  phase space ${\mathcal H}$ will be 
    costrained by the finite number of constraints.     
    What are the reasons to consider the constructed theory to be string related theory? Let the set ${\mathcal V}_{str}\subset {\mathcal H}_{str}$
     be the surface of first type constraints (\ref{con_x})  in the standard string   phase space ${\mathcal H}_{str}$
           and the  set ${\mathcal V}\subset {\mathcal H}$ --  first type constraint surface in the constructed phase
     space ${\mathcal H}$. We will have the following one-to-one correspondence $\leftrightarrow$:
     $$ \left({\mathcal V}_{str}/{\sf G}_0\right)   \longleftrightarrow  {\mathcal V}\subset {\mathcal H}\,.    $$
     Moreover the correspondence $\leftrightarrow$ will be constructed so that the momentum and the angular moment for the defined dynamical system 
     will be      equal to the N\"oether momentum and the angular moment for string. 
     Thus both physical degrees of  freedom and the dynamical invariants will coincide for the string and the constructed dynamical system 
     on the classical level.  This fact makes it possible  to interpret the constructed dynamical system  as some finite extended object on a plane.
      We emphasize that there is no canonical transformation which connects
    the  phase space ${\mathcal H}_{str}$  and the  phase space ${\mathcal H}$. 
    As the result, the quantum theory ( that is constructed in     section 4  with the help of boson-fermion correspondence method)
     differs from the standard quantum theory for strings.
               
  Let us execute the programm outlined above. Firstly we
 define  a pair of light-like and scale-invariant vectors in space $E_{1,3}$:
 \begin{equation}
 \label{e_pm}
 {\bf e}_{\pm}(\xi_\pm) =  \pm{1\over\varkappa} \,\partial_{\pm}{\bf X}(\xi_\pm)\,,
  \end{equation}
 where $\varkappa$ is an arbitrary positive constant. If the vectors ${\bf X}\in E_{1,3}$ are transformed as
 ${\bf X} \to \widetilde{\bf X}  = \lambda{\bf X} $, the constant $\varkappa$ is transformed as
 $\varkappa  \to \widetilde\varkappa =\lambda\varkappa$. Thus we separate out the scale-transformed mode by the introduction of 
 the variable $\varkappa$ and the projective vectors ${\bf e}_{\pm}$. As the action (\ref{action}) describes the scale-invariant theory,
  we consider this step to be  justified here.
  
  Secondly we  define   a pair of orthonormal bases \cite{TMF07}
 ${\bf e}_{\nu\pm}(\xi_\pm)$ that satisfy the conditions
 ${\bf  e}_{\pm}=\left({\bf e_{0\pm}} \mp {\bf e_{3\pm}} \right)/2$.
 Instead of vectors  ${\bf e}_{\nu\pm}$ we can consider  the vector-matrices  ${\bf\hat E_{\pm}}$:
 \begin{equation}
{\bf\hat E_{\pm}}= {\bf e_0}_\pm {\boldsymbol{1_2}} - \sum_{i=1}^3{\bf e_i}_{\pm}{\boldsymbol{\sigma_i}}\,,
     \label{matr_E}
\end{equation}
these matrices are more convenient here.
 We require that all the other elements of the matrix  ${\bf\hat E_+}$  (${\bf\hat E_-}$)depend on the variable $\xi_+$ ($\xi_-$ )
 only just as the vector ${\bf  e}_{+}$ (${\bf  e}_{-}$) does.
     It is clear that the definition of the bases ${\bf e}_{\nu\pm}(\xi_\pm)$ 
has three - parameter arbitrariness in each point $(\xi^0,\xi^1)$; we intend to return to this question later. 
  The principal object of our approach   is the $SL(2,C)$ - valued field
$K(\xi^0,\xi^1)$  which  is defined as follows:
\begin{equation}
{\bf\hat E_+} = K{\bf\hat E_-}K^{+}\,.
     \label{field_K}
\end{equation}
To make the reduction to D = 1+2 space-time  we  require
 that matrix $K \in SL(2,R)$. This requirement means that
$${\bf e_{2+}}(\xi_+) = {\bf e_{2-}}(\xi_-) = {\displaystyle\sf\bf b}_2\,,$$
where ${\displaystyle\sf\bf b}_2$ is a constant spatial vector.  Thus  the reduced  space-time is
any space $E_{1,2} \perp {\bf b}_2$. All these spaces are equivalent here.

 In accordance with the definition,  the matrix field $K(\xi^0,\xi^1)$  satisfies  to (special) WZWN - equation
\begin{equation}
\partial_+\left(K^{-1}\partial_-K\right)=0\,.
     \label{eq_K}
\end{equation}

Let us define the real functions $\varphi(\xi^0,\xi^1)$ and
$\alpha_\pm(\xi^0,\xi^1)$ by means of Gauss decomposition for the
matrix $K(\xi^0,\xi^1)$:
\begin{equation}
K = \left(\begin{matrix}
1&0\\
{-\alpha_+}&1
\end{matrix}\right)
\left(\begin{matrix}\exp(-{\varphi}/{2})&0\\
0&\exp({\varphi /2})\end{matrix}\right)
\left(\begin{matrix}1&\alpha_-\\
 0&1\end{matrix}\right)\,.
     \label{gauss}
\end{equation}
In general, these functions   are singular because the decomposition
(\ref{gauss}) is not defined for the points where the principal
minor $K_{11}$ vanishes. Let us introduce  regular functions
$\rho_{\pm}=(\partial_{\pm}\alpha_{\mp})\,{\exp}(-\varphi)$.
As the
consequence of the equality (\ref{eq_K})we will get  the following PDE
- system:
\begin{subequations}
\label{TL_rho}
     \begin{eqnarray}
     \label{1_sys}
     \partial_+\partial_-\varphi &=& 2\rho_+\rho_-\exp\varphi,\\
     \label{2_sys}
     \partial_\pm\rho_\mp &=& 0,\\
     \label{3_sys}
     \partial_\pm\alpha_\mp &=& \rho_\pm\exp\varphi.
     \end{eqnarray}
\end{subequations}
This system is the direct consequence of the equations (\ref{eq_x}) and constraints (\ref{con_x}) for the defined variables  
$\varphi(\xi^0,\xi^1)$ and $\alpha_\pm(\xi^0,\xi^1)$.
For the first time this PDE - system was considered  in the work \cite{PogrTal}, where the new integrable  field model was suggested 
in $2D$ space-time. 
 The introduction of the function $\varphi$ and the functions $\rho_{\pm}$  as the world-sheet parameters 
 is justified by the following formulae for the first  $({\bf I})$
and the second $({\bf II})$ forms of the world-sheet:
$$ {\bf I} =  -\frac{\varkappa^2}{2}\,{\rm e}^{-\varphi}d\xi_+d\xi_-\,,\qquad
{\bf II} = \varkappa [\rho_+ d\xi^{2}_+   -  \rho_- d\xi^{2}_- ]  \,. $$

The standard method of geometrical description of a string
\cite{BarNes} uses the equations (\ref{1_sys}) and (\ref{2_sys})
 deduced from the Gauss and Peterson-Kodazzi equations. In
the standard approach the inequalities $\rho_\mp > 0$ are fulfilled.
In this case the conformal transformations
\begin{equation}
\label{conform}
\xi_\pm \longrightarrow \widetilde\xi_\pm = A_\pm(\xi_\pm)\,,\qquad A^{\prime}\not= 0\,,
\end{equation}
allow to reduce the equation (\ref{1_sys}) to the Liouville equation; the form  ${\bf I}$  will be
the only fundamental geometrical object here. We are considering the enlarged class of the world-sheets for which   real functions $\rho_\pm$ 
will be arbitrary differentiable functions. For example, the identity $\rho(\xi) \equiv 0$ should be fulfilled on any interval
 $[a,b]\subset [0,\pi]$.
We must  emphasize that in this case there are no  transformations (\ref{conform})  that reduce
the equation (\ref{1_sys}) to the Liouville equation globally.
The group ${\sf G}$ of the  system (\ref{TL_rho}) invariancy
is much  wider then  the group (\ref{conform}).
Indeed, let the functions $\varphi(\xi_+,\xi_-)$,  $\rho_\pm(\xi_\pm)$
and $\alpha_\pm(\xi_+,\xi_-)$
 be solutions for the system (\ref{TL_rho}). Then the transformation
\begin{equation}
     (\varphi, \rho_\pm,             \alpha_\pm)\longrightarrow
(\tilde\varphi, \tilde\rho_\pm, \tilde\alpha_\pm),
     \label{group_G}
     \end{equation}
     gives the new solution for the system (\ref{TL_rho}) if
     \begin{eqnarray}
  \tilde\varphi(\xi_+,\xi_-)&=&\varphi(A_+(\xi_+),A_-(\xi_-))+
     f_+(\xi_+)+f_-(\xi_-),\nonumber\\
  \tilde\rho_\pm(\xi_\pm)&=&
\rho(A_\pm(\xi_\pm))A_\pm^{\prime}(\xi_\pm)\exp{(-f_\pm(\xi_\pm))},
     \nonumber\\
\tilde\alpha_\pm(\xi_+,\xi_-)&=&
\alpha_\pm(A_+(\xi_+),A_-(\xi_-))\exp{(f_\pm(\xi_\pm))}
+g_\pm(\xi_\pm).\nonumber
     \end{eqnarray}
     for arbitrary real functions $f_\pm(\xi)$,  $g_\pm(\xi)$ and such real functions
    $A_\pm(\xi)$ where the conditions $A_-^{\prime}A_+^{\prime}\not= 0$ are fulfilled.
From the geometrical point of view,
 two kinds of the transformations  (\ref{group_G}) exist. The first kind  corresponds to the conformal
reparametrizations of the same world-sheet. The equalities
\begin{equation}
\label{same}
f_\pm(\xi) = -\ln A_\pm^\prime (\xi)
\end{equation}
extract these transformations from the group ${\sf G}$. The second kind is all the other transformations
which connect  different world-sheets.
%  The  solution for system (\ref{TL_rho}) for which $\rho_\pm(\xi_\pm)\equiv 0$ and another functions are constants
%   will be the natural ''vacuum'' for this system. We demand  $ \varphi(\xi^0,\xi^1) \ge 0$ so that value $\varkappa$ 
%   introduced above parametrizes  the ''classical vacua'' here. 
%  Thus this constant will be a dynamical variable,  not an {\it ''in-put''} constant.

\section{Factorization prosedure.}
\nopagebreak
\indent

In this section we will be investigating the orbits of the group ${\sf G}$. The results  obtained here will help us
to construct  the anyon model  -- as certain quantum system that has exitations with arbitrary spin.
This effect is due to the property of the  group $SO(2)$ only; that is why both relativistic and non-relativistic models are interesting.
Our consideration started with relativistic objects, but in the next section we are going to reduce our theory to non-relativistic case. 
   The relativistic case will be considered in a separate work.

Now we  continue the  investigations  of the  local properties of the objects  on interval $\xi^1 \in [0,\pi]$. 
The boundary conditions will be
taken into account later.  
Let the vectors ${\displaystyle\sf\bf b}_\mu \in E_{1,3}$ be constant vectors so
that ${\displaystyle\sf\bf b}_\mu {\displaystyle\sf\bf b}_\nu =
g_{\mu\nu}$.
 Let the vector - matrix ${\bf\hat E_{0}}=
{\displaystyle\sf\bf b}_0   {\boldsymbol{1_2}} - \sum_{i=1}^3
{\displaystyle\sf\bf b}_i  {\boldsymbol{\sigma_i}}$ correspond to
the basis ${\displaystyle\sf\bf b}_\mu$. It is clear that
\begin{equation}
\label{ETT}
{\bf\hat E}_\pm (\xi_\pm)  = T_\pm(\xi_\pm) {\bf\hat E_{0}} T^{\top}_\pm(\xi_\pm)\,,
\end{equation}
where $T_\pm(\xi) \in SL(2,R)$. 
The equality
\begin{equation}
\label{KTT}
K(\xi^0,\xi^1) = T_+(\xi_+)T^{-1}_-(\xi_-)   \,
\end{equation}
is a sequence of the formula (\ref{field_K}).
Our next step is the reconstruction of the  tangent vectors  $\partial_{\pm}{\bf X}(\xi_\pm)$   through
the matrix elements $t_{ij\pm}$ of the matrices $T_\pm$. Taking into
account  the formula (\ref{ETT}) and the definition of the matrices
${\bf\hat E}_\pm (\xi_\pm)$, we obtain the following equalities:
\begin{equation}
  \pm   \partial_{\pm}{\bf X}(\xi_\pm)
 = \frac{\varkappa}{2}\Bigl[ \left( t_{i1{\pm}}^2 + t_{i2{\pm}}^2 \right)
  \,{\displaystyle\sf\bf b}_0 -
    2\left( \,t_{i1{\pm}}{t}_{i2{\pm}}\right)\, {\displaystyle\sf\bf b}_1
   - \left(t_{i1{\pm}}^2 - t_{i2{\pm}}^2\right)\,{\displaystyle\sf\bf b}_3 \Bigr] \,,
\label{dX}
 \end{equation}
  where index  $i$  corresponds to the sign $\pm$ according to the rule     $i=\frac{3\mp1}{2}$.
To reconstruct the whorld-sheet from the derivatives $\partial_{\pm}{\bf X}$ we must add the constant vector ${\bf Z}$.

The following proposition can  be deduced directly from the definitions of the matrices $T_\pm$ and $K$:

\begin{proposition}
\label{prop_2.1}
The matrices $T_\pm$ are the solutions for the linear problems
\begin{equation}
     T^{\,\prime}_\pm(\xi)+
Q_\pm(\xi)T_\pm(\xi) = 0 \,,
     \label{spect1}
     \end{equation}
where
\begin{equation}
\label{Q_def}
Q_-(\xi^0,\xi^1) = K^{-1}\partial_-K\,,\quad Q_+ (\xi^0,\xi^1) =- (\partial_+K)K^{-1}\,.
\end{equation}
\end{proposition}

The global Lorenz transformations in our (3D) theory are the transformations
\begin{equation}
\label{E_transf}
{\bf\hat E_{0}} \longrightarrow {\widetilde{\bf\hat E}_{0}}
= {\mathcal B}{\bf\hat E_{0}}{\mathcal B}^{\top}\,,
\end{equation}
where the constant matrix ${\mathcal B} \in SL(2,R)$. It is clear that these transformations correspond to
the arbitrariness for the matrix - solution of the systems (\ref{spect1}):
\begin{equation}
\label{T_transf}
 T_\pm \longrightarrow {\widetilde T}_\pm = T_\pm {\mathcal B}^{-1}\,.
\end{equation}

 Thus the coefficients of the problems (\ref{spect1}) are local functions of the introduced
  variables $\varphi$, $\rho_\pm$  and $\alpha_\pm$.   These coefficients are relativistic invariants.
  For example,   the equalities
\begin{equation}
\label{Q_rho}
 Q_{12+} = -\rho_+\,,\qquad     Q_{21-} = -\rho_-\,,
\end{equation}
 will be important for our subsequent considerations.

Let  ${\sf G}_0$ be the subgroup of the group ${\sf G}$ so that $A_\pm(\xi)\equiv \xi$ for all transformations
(\ref{group_G}). Then the following proposition is true:

\begin{proposition}
\label{prop_2.2}
If the group ${\sf G}_0$ transforms the solution \{$\varphi$,
$\rho_\pm$,  $\alpha_\pm$\} of the system (\ref{TL_rho}), the
matrices  $T_\pm$ are transformed as follows:
\begin{equation}
T_\pm\longrightarrow\tilde T_\pm={G}_\pm^{-1}T_\pm\,,
\label{T_trasf}
\end{equation}
where
$${G}_+=
\begin{pmatrix}
{\rm e}^{{f_+}/{2}}&~&0\\
~&~&~\\
g_+{\rm e}^{-{f_+}/{2}}&~&{\rm e}^{-{f_+}/{2}}
\end{pmatrix}\,,
\qquad
{G}_- =
\begin{pmatrix}
{\rm e}^{-{f_-}/{2}}&~& g_-{\rm e}^{-{f_-}/{2}}\\
~&~&~\\
0&~&{\rm e}^{{f_-}/{2}}
\end{pmatrix}\, .$$
~~\\
\end{proposition}

{\bf Proof.}
The proof is a direct consequence of the formulae (\ref{gauss}),
(\ref{KTT}) and an explicit form for the transformations
(\ref{group_G}). 
~~

Let us take into account the boundary conditions for the  field ${\bf X}(\xi^0,\xi^1)$.
The standard analysis   leads to equalities
\begin{equation}
\label{e_2pi}
  {\bf  e}_{+}(\xi) = {\bf  e}_{-}(-\xi)\,,\qquad {\bf  e}_{+}(\pi+\xi) = {\bf  e}_{-}(\pi-\xi)\,. 
  \end{equation}
 These equalities mean that we can consider $2\pi$-periodical vector field ${\bf e}(\xi) \equiv {\bf e}_{+}(\xi)$ 
 which is defined for all real $\xi$ instead of the fields ${\bf  e}_{+}$ and ${\bf  e}_{-}$ for $\xi\in [0,\pi]$.
 The function ${\bf  e}_{+}$  (or  the element $({\bf\hat E}_+)_{11}$)      and the function ${\bf  e}_{-}$ (or the element
  $({\bf\hat E}_-)_{22}$)  are constrained by the conditions (\ref{e_2pi}).
   We extend these constraints  on all elements of the matrices ${\bf\hat E_{\pm}}(\xi)$.
  Thus the   matrices       ${\bf\hat E}_{\pm}(\xi) $ will be $2\pi$-periodical  matrices on real axis and 
  ${\bf\hat E_+}(\xi) = K_0{\bf\hat E_-}(-\xi)K_0^{+}$, where $K_0 =i {\boldsymbol{\sigma_2}}$.
    Consequently, the equalities
 \begin{equation}
 \label{T_perod}
 T_-(\xi) = - K_0T_+(-\xi)\,,\qquad
 T_+(\xi+2\pi) = \pm T_+(\xi)\,, \qquad 
 \end{equation}
will be true. Further we will be considering the matrix $T(\xi)\equiv T_+(\xi)$ only. 
It is clear that $T(\xi) = T_0(\xi)B$, where the constant matrix $B\in SL(2,R)$ and the matrix $T_0(\xi)$ satisfies the boundary
condition  $T_0(0) = I_2$.  The elements of the  matrix $T_0(\xi)$ will be single-valued functions from the coefficients $Q_{ij}$ i.e. the functions
 $\varphi$   and $\alpha_\pm$.
Thus   the parametrization (\ref{param1}) has been realized.

The formulae (\ref{KTT}) and     (\ref{T_perod}) allow us to continue the functions $\varphi(\cdot,\xi^1)$, 
$\alpha_\pm(\cdot,\xi^1)$ and $\rho_\pm(\xi^1)$ on all real axis.
  For example, $\rho_\pm(\xi_\pm) = \rho(\pm\xi_\pm)$, where $\rho(\xi)$ will be a $2\pi$-periodical differentiable function.

Going back to the group (\ref{group_G}), we can consider $2\pi$-periodical function $f(\xi)$  instead of the functions
$f_\pm$ which are connected by formulae $f_\pm(\xi_\pm) = f(\pm\xi_\pm)$, similar statement  will be true for the functions $g_\pm(\xi)$. 
 We can always demand that
  \begin{equation}
\label{zero_f}
\int\limits_0^{2\pi}f(\xi)d\xi =0\,.
\end{equation}
or redefine the constant $\varkappa$ if the eq. (\ref{zero_f}) is not true.
Next, we must restrict the set of the functions $A_\pm(\xi)$ by the condition
$$ A_\pm(\xi_\pm) = \pm A(\pm\xi_\pm)\,, \qquad  A(\xi+2\pi) = A(\xi) +  2\pi\,,$$
where $A^\prime(\xi)\not= 0$. 
Obviously, we can expand the action of the group ${\sf G}$ on $2\pi$-periodical  matrix $T(\xi)$. 
 Let ${\sf G}_0[T]$ denote the orbit of the
group ${\sf G}_0$ for matrix $T(\xi)$. Then the following
proposition  will be fulfilled.

\begin{proposition}
\label{prop_2.3}
There exists the unique $SO(2)$  matrix ${\mathcal U} \in {\sf G}_0[T]$ 
solving the $2\pi$-periodical  linear problem
\begin{equation}
{\mathcal U}^{\,\prime}(\xi)   +  {Q}(\xi)   {\mathcal U}(\xi) = 0\,,
\label{spect2}
\end{equation}
where    $Q(\xi) =
-{\rho}(\xi)\boldsymbol{\sigma_+} + {\rho}(\xi)\boldsymbol{\sigma_-}$.
\end{proposition}

{\bf Proof.}
Indeed, let us consider the Iwasawa decomposition for the matrix
$T(\xi)$ such that $T={\mathcal E N U}$ where the
matrix ${\mathcal E}$ is a diagonal matrix with positive
elements,  ${\mathcal N}$  is a lower 
triangular matrix and ${\mathcal U}\in SO(2)$. 
The statement of Proposition \ref{prop_2.3} is a sequence of the unique existence of Iwasawa decomposition for any matrix $T \in
SL(2,R)$,  rule (\ref{T_trasf}) for the matrix $T=T_+$ transformation and corresponing rule for the matrix $Q$.
~~~
~~
We use the same characters for  matrices $Q$ (the same concerns  the coefficients
$\rho$) both in the linear problem (\ref{spect1})
 and in the linear problem (\ref{spect2}); we hope that these notations won't lead to any ambiguities.

  The group ${\sf G}_0$ can be decomposed into two kinds of the  special  transformations:
\begin{eqnarray}
\label{G_A}
\alpha_\pm & \to & \alpha_\pm+g_\pm \,, \\
\label{G_B}
   \varphi & \to & \varphi + f_+ +f_-\,,\quad \rho_\pm \to \rho_\pm e^{-f_\pm}\,,\quad
   \alpha_\pm  \to   \alpha_\pm e^{f_\pm} \,. 
   \end{eqnarray}

 The following proposition is true.

\begin{proposition}
\label{prop_2.4}
The transformation (\ref{G_A}) does not change the world-sheet; the
transformation (\ref{G_B}) transforms the world-sheet to the other world-sheet such
that
\begin{eqnarray}
\label{Weyl}
 {\bf I} & \longrightarrow & \widetilde{\bf I} = {\bf I}\exp[-f_+ - f_-] \,,\\[3mm]
\label{form_II}
{\bf II}  & \longrightarrow &  \widetilde{\bf II}
= \varkappa [\rho_+ e^{-f_+}d\xi^{2}_+   -  \rho_- e^{-f_-}d\xi^{2}_- ]\,.
\end{eqnarray}
\end{proposition}

{\bf Proof.}
 The proof is a sequense of the explicit formulae (\ref{dX}) for tangent vectors $\partial_{\pm}X(\xi_\pm)$,
the explicit formulae for the forms {\bf I} and {\bf II} and rules
(\ref{T_trasf}) for transformations of the matrix elements
$t_{ij\pm}$. Note that the existence of the
transformations (\ref{G_A}), which do not change the world-sheet, is the
consequence of the arbitrariness in the definition of matrices
${\bf\hat E}_\pm (\xi_\pm)$.
~~

Let us consider  the set    of  world-sheets ${\mathcal X}$
 introduced in the beginning of the paper. It will be recalled that space-time symmetry group here is the $3D$  
 Poincar\'e group  $E(1,2)$.
The object of our subsequent investigations is the factor-set ${\mathcal X}/{\sf G}_0$ only.
Let us investigate the parametrization of the corresponding cosets.
We are going to construct the parameters that can be  separated  into two sets. 
The first set will contain the finite number of ''external''  variables 
  that  parametrize certain   space-time symmetry group in some way.
  The second set will be invariant under this group (the ''internal'' variables). 
For the set ${\mathcal X}$, for example, the  ''external''  variables are  the constant vector
${\bf Z}\in E_{1,2}$,  the    matrix $B\in SL(2,R)$ (see (\ref{param1})) which parametrize  the group $E(1,2)$ locally,
and the quantity $\varkappa$.

Taking into account   the Proposition 2.3 we can select the representatives in every coset so that
$T_\pm = {\mathcal U}_\pm \in SO(2)$. Thus we have
\begin{equation}
\label{UU}
{\mathcal U}(\xi) \equiv {\mathcal U}_+(\xi) = {\mathcal U}_{0}(\xi)U(\beta)\,,
\qquad
{\mathcal U}_{0}(0) = 1_2\,, \qquad {\mathcal U}(\beta) = \left(\begin{matrix}
   \cos\beta &  \sin\beta\\
- \sin\beta& \cos\beta
\end{matrix}\right)\,.  
\end{equation}

   Because of the boundary conditions for the matrix ${\mathcal U}_{0}(\xi)$, the
one-to-one correspondence $\rho(\xi) \leftrightarrow {\mathcal
U}_{0}(\xi)$ exists.      It is easy to see that
 \begin{equation}
\label{matr_U}
 {\mathcal U}_{0}(\xi) =
 \left(\begin{matrix}
   \cos I(\xi)&  \sin I(\xi)\\
- \sin I(\xi)& \cos I(\xi)
\end{matrix}\right)\,,
\qquad I(\xi) = \int_{0}^\xi\rho(\eta)d\eta\,.
  \end{equation}
 In accordance with the second  formulae  (\ref{T_perod}) the matrix ${\mathcal U}_{0}(\xi)$ must be  (anti)periodical.
 This fact  means that the condition
    \begin{equation}
  \label{topol}
  \int_{0}^{2\pi}\rho(\eta)d\eta = \pi n\,,\qquad n= 0,\pm 1, \pm 2. \dots
  \end{equation}
  must be fulfilled.    
  Tangent vectors $\partial_{\pm}{\bf X}(\xi_\pm)$ are defined through $2\pi$-periodical vector-function  ${\bf e}(\xi)$
 as follows:
  \begin{equation}
\label{tangent}
  \pm \partial_{\pm}{\bf X}(\xi_\pm)  =  {\varkappa}{\bf e}(\pm\xi_\pm)\,,
  \end{equation}
  where
    $$   {\bf e}(\xi)=  {1\over 2} \Bigl[{\displaystyle\sf\bf b}_0 - \sin(2I(\xi) + 2\beta) {\displaystyle\sf\bf b}_1
  - \cos (2I(\xi) +2\beta){\displaystyle\sf\bf b}_3\Bigr]  \,.$$
 It is clear that
  $ X_0(\xi^0,\xi^1) =  \varkappa\xi^0 {\displaystyle\sf\bf b}_0 +  Z_0$
  for our gauge.   To reconstruct the spatial coordinates ${X}_j(\xi^0,\xi^1)$  ($j=1,3$) of the world-sheet  through the derivatives,  
  we must introduce a two-dimensional vector with   components $Z_1$ and $Z_3$.  Thus we have the following one-to-one correspondence:
  \begin{equation}
\label{corresp}
 \Bigl(   X_1(\xi^0,\xi^1),\, X_3(\xi^0,\xi^1) \Bigr) \longleftrightarrow
 \Bigl(\varkappa;\, Z_1, Z_3,\,  \beta;\,  \rho(\xi) \Bigr)\,.
 \end{equation}
 The  variables  $\rho(\xi)$ and $\varkappa$ will be invariant under the group  $E(2)\times {\mathcal T_0}$, 
where $E(2)$ is the group of the motions for the spatial plane  $E_2\perp {\displaystyle\sf\bf b}_2$ and
${\mathcal T_0}$  is the group of  time shifts.
  The  variables $( Z_1, Z_3,  \beta)$ are transformed under space translations and space rotations in obvious manner.

 Thus the  following proposition will be true.
\begin{proposition}
\label{prop_2.5}
A space-time symmetry group for set ${\mathcal X}/{\sf G}_0$  will be the group  $E(2)\times {\mathcal T_0}$. 
\end{proposition}

Where were the Lorentz boosts  lost?
  It appears that  two operations are non-commutative: the 
   boost in the space-time $E_{1,2}$ 
and the selection of the gauge $T(\xi)\equiv {\mathcal U}(\xi)$.
 Thus the Lorentz  boosts transform  the functions $\rho(\xi)$ one through the other 
  and will be ''internal'' transformations here.

In the context of the factorization procedure  defined above,
  we can write the principal minor $K_{11}$ of the matrix $K(\xi^0,\xi^1)$
  as the function of the quantity $\rho$. To do it we must extract the element $K_{11}$
 from the formula (\ref{gauss}). The result is as follows:
 \begin{equation}
\label{phi_rho}
 \exp[-\varphi(\xi^0,\xi^1)] = \sin^2 \int_{-\xi_-}^{\xi_+}\rho(\eta)d\eta \,,\qquad \xi_\pm = \xi^1\pm\xi^0\,.
 \end{equation}
 This equality can be considered as the  geometrical gauge condition for our theory.
 It must be emphasized that the arbitrariness (\ref{conform}) has not been fixed anywhere.

 \section{Dynamical system.}
\indent

Let us write the formulae for  N\"oether invariants of the action
(\ref{action}):
$$P_\mu = \gamma \int_{0}^{\pi} \partial_0 X_\mu\, d\xi^1\,,\qquad
M_{\mu\nu} = \gamma \int_{0}^{\pi} \left(\partial_0 X_\mu
X_\nu  -  \partial_0 X_\nu  X_\mu\right) d\xi^1\,. $$

The formulae (\ref{tangent}) make it possible to calculate the components $P_\mu$  through the
variables  $\rho(\xi)$,      $\beta$, and  $\varkappa$; for example, the string energy $P_0 = \pi\gamma\varkappa$. 
As it has been proved above,  space-time symmetry group of our  system will be the group  $E(2)\times {\mathcal T_0}$; 
and that is why we will use the formulae  for  N\"oether invariants for spatial indices only. 
The following expressions can be deduced for the  quantities
${\bf P}^2 = P^2_1 + P^2_3$ and ${\sf S} = M_{13} - Z_1 P_3 - Z_3 P_1\,$:
\begin{equation}
\label{mass}
{\bf P}^2 =\pi^2{\gamma^2\varkappa^2}{F_P[\rho]} \,,
\end{equation} 

\begin{equation}
\label{spin}
{\sf S} = \frac{\pi\gamma\varkappa^2}{2}{F_S[\rho]}\,,
\end{equation} 

where 
\begin{eqnarray}
{F_P[\rho]} & = & \frac{1}{4\pi^2}  \int\limits_0^{2\pi}\int\limits_0^{2\pi}d\xi d{\overline\xi}
\cos\Bigl(2\int_{\overline\xi}^\xi \rho(\eta)d\eta\Bigr)\,,\nonumber\\
{F_s[\rho]} & = & \frac{1}{2\pi}\int\limits_0^{2\pi}\int\limits_0^{2\pi}d\xi d{\overline\xi}\,
W(\xi - {\overline\xi})\sin\Bigl(2\int_{\overline\xi}^\xi \rho(\eta)d\eta\Bigr)\,,\nonumber\\
W(\xi)& = &\frac{1}{2\pi i}\sum_{n\not=0} \frac{1}{n}\, e^{-in\xi} = 
\frac{\xi}{2\pi} - \Bigl[\frac{\xi}{2\pi}\Bigr] - \frac{1}{2}\,.\nonumber 
\end{eqnarray}

\begin{proposition}
\label{prop_3.1}
The quantities $P_1$, $P_3$, ${\sf S}$ and $\rho(\xi)$  are constrained by the
following condition:
\begin{equation}
\label{constr_1}
 2\pi\gamma{\sf S} F_P[\rho] = {\bf P}^2 F_s[\rho]\,.
\end{equation}
\end{proposition}

{\bf Proof.}
The proof is the exclusion of the variable $\varkappa$ from the formulae   (\ref{mass}) and (\ref{spin}).
~~

The dynamical system corresponding to the set ${\mathcal X}/{\sf G}_0$ is defined as follows.
 \begin{itemize}
 \item Space-time $E_{1,2}$ is reduced to ''space and time'' $E_2$ and $R_1$. The corresponding group of space-time symmetry 
 is reduced to the group $E(2)\times {\mathcal T_0}$.
 Thus there are no world sheets (as  geometrical objects) 
from this moment but the moving planar curves still exist.
 \item We define the constants $(P_1, P_3, {\sf S})$  as the  dynamical variables in our theory instead the variables 
 $\beta$ and $\varkappa\in (0,\infty)$. The corresponding set of the variables 
$({ \rho(\xi)\, ;  P_1, P_3; B_1, B_3; {\sf S} })$ we note as ${\mathcal W}$.
Since we introduced three variables instead of two,
the condition (\ref{constr_1}) must be imposed as  the constraint; the symbol ${\mathcal V}$ denotes the corresponding surface.
Then the following one-to-one correspondence is fulfilled:
 \begin{equation}
  \label{one_one}
    {\mathcal X}/{\sf G}_0  \longleftrightarrow  {\mathcal V} \subset {\mathcal W}\,.
  \end{equation}
\item Supposing the variables $(P_1, P_3, {\sf S})$ and $\rho$ are independent we  must close the domain $(0,\infty)$ 
for the constant $\varkappa$ by adding the boundary points $\varkappa = 0$ and $\varkappa = \infty$. Indeed,
in accordance with our initial supposition the constant $\varkappa$  is a non-zero finite constant; for this domain
the identity $\rho\equiv 0$ leads to the equalities $|{\bf P}| =0$, ${\sf S}=0$.
 Of couse there are no strings that correspond to the points $\varkappa = 0$ and $\varkappa = \infty$. 
 \item We extend the group  $E(2)\times {\mathcal T_0}$  to  Galilei  group ${\mathcal G}_2$. Indeed,
 the transformation 
  $$  {\bf P} \to \widetilde{\bf P} = {\bf P} + c{\bf v}\,, \qquad 
   {\bf v} = v_1 {\displaystyle\sf\bf b}_3  + v_3  {\displaystyle\sf\bf b}_3\,\quad (c,v_j = const)  $$
   defines Galilei boosts on the set of (independent) cordinates $( \rho(\xi^1+\xi^0)\,;$ ~~ $ P_1, P_3; Z_1, Z_3; {\sf S})$
   and is quite natural here. 
  \item  The central extension $\widetilde{\mathcal G}_2$  will be considered  
instead  of ${\mathcal G}_2$;   this step allows us to introduce an additional ''in-put'' parameter $m_0$ as a central charge and quantize 
theory\,\footnote{We consider the one-parameter extension only.}.
  \item We use the variables
$$  B_j = m_0\left( Z_j - \frac{\xi^0}{\gamma}\,P_j\right)\,, \qquad j=1,3 $$
instead of the variables $Z_j$.
\end{itemize}

After the reduction to the nonrelativistic case the following problem appears: what function   will be the energy of constructed dynamical system?
   There are three Cazimir functions  for central extended Galilei algebra:
    $$ {\hat C}_1 = m_0 {\hat I}\,,\quad 
  {\hat C}_2 =  \Bigl[ {\hat M}_{13} - {\hat B}_1 {\hat P}_3 - {\hat B}_3 {\hat P}_1 \Bigr]^2\,, 
  \quad {\hat C}_3 = \hat H -  ({1}/{2m_0}){\hat {\bf P}}^2   \,,$$
       where         ${\hat I}$ is a unit operator,  quantities  ${\hat M_{13}}$,   $\hat H$,  ${\hat P}_i$  
        and  ${\hat B}_i$ --  generators of rotations, time and space translations and Galilei boosts correspondently.
        It is well-known that Cazimir function    $C_3 $  is interpreted as the internal energy of a ''particle'' (i.e. of our dynamical system). 
        Thus the  definition of the full energy  as  the function
 \begin{equation}
  \label{energy_1}
{\sf E} = \frac{{\bf P}^2}{2m_0} + h[\rho]\,,
\end{equation}
   where  $h[\rho]$ is the hamiltonian for ''internal'' variable $\rho(\xi^1+\xi^0)$, will be quite natural.

Let us define the hamiltonian structure in our theory  as follows:
\begin{itemize}
\item the phase space ${\mathcal H} = \overline{\mathcal W}$ with fundamental 
coordinates\,\footnote{closure in the weak topology that is defined by the function $\varkappa =\varkappa[\rho(\xi);{\sf S}]$}
$$ \Bigl( \rho(\xi)\,;\,P_1, P_3,\, B_1, B_3; \, {\sf S}\Bigr)\,; $$
\item
Poisson brackets
\begin{eqnarray}
\label{br_rho}
\{\rho(\xi),  \rho(\eta)\} & = &  -\frac{1}{4}\, \delta^\prime(\xi - \eta)\,, \\[3mm]
\{P_i,B_j\} &=& m_0\delta_{ij} \,
\end{eqnarray}
(other possible brackets  equal  zero);
\item
constraints (\ref{topol}) and (\ref{constr_1}) (the constraint surface will be the set $\overline{\mathcal V}$);
\item
hamiltonian
$$ H =   \frac{{\bf P}^2}{2m_0}  + 2\int_{0}^{2\pi} \rho^{\,2}(\xi) \,d\xi +
l(\xi^0)\Phi\,,$$ where the function $l$ is a lagrange multiplier.
\end{itemize}

 The phase space will be as
follows:
$$ {\mathcal H} = {\mathcal H}_\rho \times  {\mathcal H}_2\,,$$
where  ${\mathcal H}_\rho $ is the phase space of internal degrees of  freedom 
(it is parametrized by the function $\rho(\xi)$) and ${\mathcal H}_2$ is the phase
space of a free particle on a plane $E_2$. The model is non-trivial
because of the constraint (\ref{constr_1}) that entangles the
internal and external  variables.
 Topological constraint (\ref{topol}) selects the symplectic sheets in the space ${\mathcal H}_\rho $.

 \section{Quantization}
 \indent
 
 The theory considered above 
 leads to the following natural structure of the Hilbert space of the quantum states: 
  $$ {\bf H} = \int_S\,{\bf H}_S\,, \qquad
    {\bf H}_S = {\bf H}_2 \times {\bf H}_\psi\,.$$
 The space ${\bf H}_2 \approx L^2 ({\sf R}_2)$ is the Hilbert space for  a free
 non-relativistic particle with internal moment $S$ on a plane;  the space
 ${\bf H}_\psi$  is the Fock space of   the ''internal degrees of  freedom'' $\rho(\xi)$.
The constraint  (\ref{constr_1}) leads to the equation for physical states  $|\psi_s\rangle \in {\bf H}_S$: 
\begin{equation}
\label{spectrum_2}
\Bigl(\gamma{\sf S}{\hat I}_2\otimes {\widehat F}_P - 
 \widehat{\boldsymbol{P}}^2\otimes \widehat{F}_S\Bigr)|\psi_s\rangle  =  0\,.   
\end{equation}
 The stationary Schr\"odinger equation
  \begin{equation}
  {\widehat H}|\psi_s\rangle \equiv
 \Bigl((1/2m_0) \widehat{\boldsymbol{P}}^2\otimes{\hat I}_\psi   + {\hat I}_2\otimes {\hat h}   \Bigr)|\psi_s\rangle  = 
 {\sf E}|\psi_s\rangle\,
 \label{Schr}
  \end{equation} 
    defines the energy of our system   together with the  equation (\ref{spectrum_2}). The following notations are used here:
  symbol  ${\widehat\dots}$ denotes the  quantized functions in the corresponding space so that
   ${\hat I}_i$ is the unit operator in the space ${\bf H}_i$ ($i=2,\psi$) and so on. 
 The states $|\psi_s\rangle \in {\bf H}_S$ that solve the system    (\ref{spectrum_2}) --  (\ref{Schr}) can be considered as 
 entangled states such that
 $$ |\psi_s\rangle = \sum_{n} a_n |f_{s,n} \rangle |\alpha_n\rangle\,,$$
where
 $ |f_{s,n} \rangle \in {\bf H}_2$  and   $|\,\alpha_n\rangle  \in {\bf H}_\psi$.
 Let the space ${\bf H}_2$ be the (framed) space $L^2 ({\sf R}_2)$ so that  $\widehat{\boldsymbol{P}}^2 = -\Delta_2$.
 Suppose that the wave functions $\langle {\bf Z}|f_{s,n} \rangle$ have a form
   $$ \langle {\bf Z}|f_{s,n} \rangle = J_{l-s}(k_{s,n}r)e^{i(l-s)\phi}\,,\quad\qquad  {\bf Z}:\,(Z_1 = r\cos\phi,\, Z_3 =r\sin\phi)\,,$$
  where the functions   $J_\nu$ are Bessel functions and number $l$ is a total  moment of the whole system. 
  Then  the real quantities  $k=k_{s,n}$ and vectors $|\,\alpha\rangle =|\,\alpha_n\rangle $  are found from the following 
 spectral problem in the space ${\bf H}_\psi$:
 
 \begin{eqnarray}
 \label{sys_1}
 \Bigl(2\pi\gamma\,{\sf S}{\widehat F}_P - k^2   \widehat{F}_S\Bigr)|\,\alpha\rangle  &=&  0\,,\\  
 \label{sys_2}
 \Bigl(\frac{{k^2}}{2m_0}    + {\hat h}   \Bigr)|\,\alpha\rangle & = &  {\sf E}|\,\alpha\rangle\,.
 \end{eqnarray}

 The suggested scheme will be formal unless we quantize the function $\rho$, define the Hylbert space ${\bf H}_\psi$ and construct the
 corresponding operators ${\widehat F}_P$, $\widehat{F}_S$ and ${\hat h}$.
 To do this   we  apply the method of the boson-fermion correspondence.
   We intend to follow  the work \cite{Pogr} where  both the rigorous investigation and the detailed historical 
  review of this  procedure were carried out.
  
  Let us define the  fermionic field $\psi(\xi^0,\xi^1) = \psi(\xi^0 + \xi^1)$ where $2\pi$-periodical operator-valued function 
   $\psi(\xi)$ is defined as follows:
   $$\psi(\xi) = \sum_{n=1}^{\infty}\,a_n^{*} e^{-in\xi}  + \sum_{n=0}^{\infty}\,b_n e^{in\xi}  \,.$$
   The fermionic operators $a_n^*$ and  $b_n^*$ will be creation operators in the Fock space ${\bf H}_\psi$ with vacuum vector
    $|\,0\rangle$; the operators $a_n$  and  $b_n$ will be the corresponding annihilation operators. Canonical anticommutation relations 
      $$[a_n^{*},a_m]_+ = \delta_{nm} \quad (n,m = 1,2,\dots)\,,\qquad
   [b_n^{*},b_m]_+ = \delta_{nm} \quad (n,m = 0,1,2,\dots)\,$$
   are carried out. As the next step we consider  the current
   $v(\xi) =:\!\psi^{*}(\xi)\psi(\xi)\!:$ where the  symbol $:\!~\!:$ denotes the fermion ordering:
  $$:\!\psi^{*}(\xi)\psi(\xi)\!: = \lim_{\eta\to\xi}\Bigl(\psi^{*}(\eta)\psi(\xi) - \langle 0|\psi^{*}(\eta)\psi(\xi)|\,0\rangle\Bigr)\,.$$
  The current $v(\xi)$ will be the well-defined  bozonic field with commutation relations:
  \begin{equation}
  \label{com_rel}
  [v(\xi),v(\eta)] = i\delta^{\prime}(\xi-\eta)\,.
  \end{equation}
  
  The charge 
  $$\Lambda =\frac{1}{2\pi}\int_0^{2\pi}v(\xi)d\xi = \sum_{n=0}^\infty b_n^*b_n - \sum_{n=1}^\infty a_n^*a_n $$
  has integer eigenvalues. We can decompose the space  ${\bf H}_\psi$ as follows
    $$ {\bf H}_\psi =   \mathop{\bigoplus}\limits_{n = -\infty}^\infty   {\bf H}_n\,, $$
  where space ${\bf H}_n$ is the eigenspace  corresponding to the eigenvalue $n$ of operator $\Lambda$. Details
  can be found in the work \cite{Pogr}.

 Here is the quantization postulate for the internal degrees of freedom -- function $\rho(\xi)$:
 $$ \rho(\xi) \longrightarrow  \hat\rho(\xi) \equiv \frac{1}{2}\, v(\xi)\,.$$ 
  
  What is the motivation for this postulate here? Don't take into account the boundary conditions (\ref{bc_x}), let us return
  to the linear problems (\ref{spect1}) and consider the functions $t_{ij\pm}(\xi_\pm)$. The following proposition will be true.
  \begin{proposition}
 \label{prop_4.1}
  The objects    $   \left(\begin{matrix} t_{i1\pm} \\ t_{i2\pm} \end{matrix}\right) $  will be the Majorana spinors in the space-time $E_{1,2}$
  for every sign $\pm$ and $i=1,2$; the objects $   \left(\begin{matrix} t_{ij+} \\ t_{ij-} \end{matrix}\right) $ will be the
  spinors in tangent plane $E_{1,1}$ for every
  $i,j=1,2$.
     \end{proposition}
  {\bf Proof.} The first
statement follows from the formulae  (\ref{E_transf}) and (\ref{T_transf}).
To prove the second statement let us fulfill the Lorentz transformation for tangent plane $E_{1,1}$:
$  \xi_\pm \to   \tilde\xi_\pm = \lambda^{\pm 1}  \xi_\pm$. The formulae (\ref{dX}) demonstrate that the quantities $t_{ij\pm}$
are transformed as $t_{ij\pm}\to \tilde t_{ij\pm}  =  \lambda^{\pm 1/2}t_{ij\pm} $. Thus the  objects 
$\left(\begin{matrix} t_{ij+} \\ t_{ij-} \end{matrix}\right) $ are transformed as  spinors in the ''space-time'' $E_{1,1}$.
  ~~
  
  The elements of the matrices $Q_\pm$ are the bilinear combinations of the  objects $t_{ij\pm}$ and $t^\prime_{ij\pm}$.
  Thus the interpretation of the quantum  function $\rho$ (it defines the elements of the matrices $Q_\pm$ as considered above) as the current
  of free $2D$ fermionic field will be quite natural in our model.

  In accordance with definition of the charge $\Lambda$, the topological constraint (\ref{topol}) is fulfilled identically for 
  our quantization. 
     Let us construct the operators  ${\widehat F}_P$, $\widehat{F}_S$. 
  To do this we use\,\footnote{after the modification for
  the periodical case and chirality ''+''} the Theorem 6.1 from the work \cite{Pogr}:
  \begin{equation}
  \label{theor_61}
  \left[\exp\Bigl(-i(\xi -\overline\xi)\Bigr) - 1\right]:\!\psi^{*}(\xi)\psi(\overline\xi)\!: =
  \vdots\exp\left(i\int_{\overline\xi}^\xi v(\eta)d\eta\right)\vdots -1\,,
   \end{equation}
  where symbol $\vdots~~\vdots$ denotes the boson ordering. 
  Taking into account formula (\ref{theor_61}) and  the classical formulae for the quantities $F_P$ and $F_S$, we find the explicit
    formulae for the operators ${\widehat F}_P$ and $\widehat{F}_S$ in the space ${\bf H}_{\psi}$: 
  
  \begin{eqnarray}
  {\widehat F}_P & =& {\rm Re}\frac{1}{4\pi^2}\int\limits_0^{2\pi}\int\limits_0^{2\pi}d\xi d{\overline\xi}
  \left[1 +   \left(\exp\Bigl(-i(\xi -\overline\xi)\Bigr) - 1\right):\!\psi^{*}(\xi)\psi(\overline\xi)\!: \right] = \nonumber\\[2mm]
  ~~~ & =&  1 - b_0^*b_0 - a_1^*a_1\,,\\[2mm]
    \widehat{F}_S & =& {\rm Im}\frac{1}{2\pi}\int\limits_0^{2\pi}\int\limits_0^{2\pi}d\xi d{\overline\xi}\, W(\xi -\overline\xi)
      \left(\exp\Bigl(-i(\xi -\overline\xi)\Bigr)-1\right):\!\psi^{*}(\xi)\psi(\overline\xi)\!:\, = \nonumber\\[2mm]
 ~~~ &=&   b_0^*b_0 - a_1^*a_1  + \sum_{k=1}^\infty\,\frac{ a_{k+1}^*a_{k+1}  -b_k^*b_k}{k(k+1)} \,,     
  \end{eqnarray}

 These simple formulae justify our approach  to the quantization of bosonic field $\rho(\xi)$. 
 The following proposition is fulfilled:

 \begin{proposition}
 \label{prop_4.2}
 { The eigenvalues of operator ${\widehat F}_P$ are the integer numbers $-1$, $0$, $1$. The eigenvalues of operator ${\widehat F}_S$
  form the everywhere dence set on interval $[-2,2]$.}
  \end{proposition} 
  
  {\bf Proof.}
  The first statement is obvious. To prove the second statement let us note that $\sum_{n=1}^\infty 1/n(n+1) =1$. It is clear that
  we can approximate every number $\beta\in(0,1)$ by sum $\sum_{n=1}^\infty \epsilon_n/n(n+1)$, where the factor $\epsilon_n$ can be $0$ or $1$. 
  We omit the detailed algorithm here. 
  ~~
  
  Let us select the physical states. It is clear that the considered object does not interact with anything. 
  That is why we must exclude any states that lead to equality $k=0$ or inequality $k^2<0$ in the system (\ref{sys_1}) - (\ref{sys_2}).
  Thus we must consider the states
   $$b_{n_1}^*\dots b_{n_k}^*a_{m_1}^*\dots a_{m_l}^*|\,0\rangle\,\qquad n_i\not= n_j\,,\quad n_j\not = 0\,,
   \qquad m_i\not= m_j\,,\quad m_j\not = 1\,.$$
  
   These states correspond to the anyon with the arbitrary spin  ${\sf S}$ that  is connected with the energy by means of the formula
      \begin{equation}
      \label{ES_formula}
       {\sf S} =  \frac{m_0}{\pi\gamma}  \Bigl({\sf E} -  \sum_{i=1}^k n_i - \sum_{i=1}^l m_i  \Bigr)
       \left(\sum_{i=1}^l\frac{1}{m_i(m_i+1)} -\sum_{i=1}^k\frac{1}{n_i(n_i+1)} \right)\,.
       \end{equation}
       
     We have $  \langle\psi_{S_1}|\psi_{S_2}\rangle \propto \delta(S_1-S_2)$ for the considered states; thus 
       $|\psi_{S}\rangle \in {\bf H}_S^\prime$ where the symbol $~^\prime$ denotes the framed Hylbert space.          
  The formula   (\ref{ES_formula}) corresponds the case $k^2>0$ for  ${\sf E} > \sum_{i=1}^k n_i + \sum_{i=1}^l m_i$ only.

 \section{Concluding remarks}
 \indent

We have constructed  here the new dynamical system  on a plane.  The phase space of the constructed dynamical system has a ''string sector'' -- 
the  set ${\mathcal V}$ which is everywhere dence on the constraint surface $\overline{\mathcal V}$; this set
 corresponds bijectively to the theory of   open string on a plane.
This fact allows us to interpret this dynamical system    as the  extended particle.
Obviously the space ${\bf H}_\psi$ is redundant for quantization of the field $\rho(\xi)$. Indeed, this  space was constructed as the
  Fock space for fermionic field $\psi(\xi)$; the current $v(\xi)$ will be invariant for the transformations
  \begin{equation}
  \label{inf_psi}
  \psi(\xi) \longrightarrow \widetilde\psi(\xi) =  \psi(\xi)\exp[i\chi(\xi)]\,. 
  \end{equation}
    In our opinion, this problem can be solved in two ways.
  The first way is to  pass from the space  ${\bf H}_\psi$ to bosonic Hilbert space $ {\bf H}_B$ which is connected with the space
    ${\bf H}_\psi$ by the formula  ${\bf H}_\psi = {\bf H}_B \times {\bf H}_0$. The space ${\bf H}_0$ will be the space for ''zero mode'' 
    operator $\Lambda$ and the operator $p$ that is canonically conjugated with the $\Lambda$. Details
  can be found in the work \cite{Pogr}. The second way is to interpret the superfluous degrees of  freedom. 
 So, the ''string sector'' corresponds to the factor-set  ${\mathcal X}/{\sf G}_0$; the superfluous degrees of  freedom
 can be  used, for example,  to quantize the orbits
  of group ${\sf G}_0 $. This possibility will be investigated in subsequent works.

 In this article we did not set  ourselves  any discussing of critical dimensions in string theory as an object. 
 From the viewpoint of our approach, this question was discussed, for example, in the work \cite{TalJPh}, where the relativistic theory of the spinning string in four-dimensional space - time was suggested.

\end{document}